\documentclass[aps,prl,twocolumn,amsmath,amssymb,showpacs,superscriptaddress]{revtex4}

\usepackage[dvips]{graphicx}
\usepackage{mathrsfs}

\begin{document}

\bibliographystyle{apsrev}

\title{Error tolerance and tradeoffs in loss- and failure-tolerant quantum computing schemes}

\author{Peter P. Rohde}
\email[]{rohde@physics.uq.edu.au}
\homepage{http://www.physics.uq.edu.au/people/rohde/}
\affiliation{Centre for Quantum Computer Technology, Department of Physics\\ University of Queensland, Brisbane, QLD 4072, Australia}

\author{Timothy C. Ralph}
\affiliation{Centre for Quantum Computer Technology, Department of Physics\\ University of Queensland, Brisbane, QLD 4072, Australia}

\author{William J. Munro}
\affiliation{Hewlett-Packard Laboratories, Filton Road, Stoke Gifford, Bristol BS34 8QZ, United Kingdom}

\frenchspacing

\date{\today}

\begin{abstract}
Qubit loss and gate failure are significant problems for the development of scalable quantum computing. Recently various schemes have been proposed for tolerating qubit loss and gate failure. These include schemes based on cluster and parity states. We show that by designing such schemes specifically to tolerate these error types we cause an exponential blow-out in depolarizing noise. We discuss several examples and propose techniques for minimizing this problem. In general this introduces a tradeoff with other undesirable effects. In some cases this is physical resource requirements, while in others it is noise rates.
\end{abstract}


\maketitle

Quantum computing holds great promise for solving computational problems intractable on classical computers. A major obstacle facing all quantum computing architectures is the introduction of errors. In particular, qubit loss and gate failure are significant problems in some architectures. Most notably this affects photonic schemes, such as linear optics quantum computing \cite{bib:KLM01,bib:Kok05}. Here these types of errors arise through the physical loss of photonic qubits, the inefficiency of photon sources and detectors, and the non-determinism of multi-qubit gates.

Recently there have been several proposals for tolerating qubit loss, including ones based on cluster \cite{bib:Varnava05} and parity \cite{bib:RalphHayes05} states. There have also been proposals for tolerating gate failure \cite{bib:BarrettKok05,bib:Duan05}. These schemes achieve loss/failure tolerance through redundant encoding. This allows multiple attempts at performing measurement or gate operations, suppressing loss/failure rates. However, redundancy also introduces additional opportunities for other types of noise to be introduced, increasing effective error rates. We demonstrate that in general this results in an exponential blow-out in depolarizing noise.

In a variety of contexts this can be a serious problem. When embedded into a fault tolerant quantum computing architecture it could strongly reduce the effective fault tolerant threshold. In a loss-tolerant quantum memory it could quickly reduce the memory to a dephasing (i.e. classical) channel. In the context of state preparation strategies, which have applications beyond quantum computing, it could result in the preparation of highly mixed states.

We go on to show that in general these problems can be significantly reduced with appropriate modifications to the schemes. However, doing so introduces a tradeoff between loss/failure tolerance and other undesirable effects -- in some cases physical resource requirements, and in others different error types. This fundamentally limits the loss/failure tolerance of these schemes.

We begin by introducing the notion of error teleportation, sometimes referred to as error propagation. This occurs when qubits in an entangled state are subject to noise and subsequently measured, causing the noise to be teleported onto the other qubits. Error teleportation is the physical basis for undesirable error scaling in the schemes we discuss. We then apply this principle to two examples: a gate-failure-tolerant state preparation scheme, and a loss-tolerant quantum computing scheme. Both these examples rely on the cluster state model for quantum computing. We do not review cluster states here and suggest the unfamiliar reader refer to Refs.~\cite{bib:Nielsen06,bib:Raussendorf01,bib:Raussendorf03}.

The first example we consider is a gate-failure-tolerant scheme for constructing cluster states, which are a resource for universal quantum computation. We show that while this scheme is tolerant against gate failure, it exponentially magnifies the effects of depolarizing noise. We describe a modification to the scheme which minimizes this problem. However, this introduces a tradeoff between failure tolerance and physical resource requirements. We also provide a more general discussion of state preparation strategies in the context of error propagation.

The second example is a scheme for tolerating qubit loss in the cluster state model for quantum computing. We demonstrate that the loss tolerance of this scheme also causes an exponential blow-out in error rates. Again we suggest a modification to the protocol to overcome this problem. Doing so presents a tradeoff between loss- and error-tolerance and significantly reduces the otherwise very high loss-tolerance promised by the scheme.

Our results suggest that specialized loss/failure tolerant protocols will be limited to dealing with comparatively modest levels of loss/failure, and in realistic scenarios will be unable to achieve their otherwise very high thresholds.

We now describe the principle of error teleportation. Consider an $n$ qubit, maximally entangled state. If any single qubit is measured, the state of the remaining qubits is projected into a smaller pure state. Now suppose the measured qubit was first depolarized. This decorrelates the qubit from the remaining qubits and the measurement outcome will have no correlation with the remaining state. Thus, depolarization followed by measurement is equivalent to tracing a qubit out. This leaves the remaining state in the completely mixed state -- the noise has been teleported onto the other qubits.

Next we consider how error teleportation scales in a situation where multiple qubits are measured. We assume an independent depolarizing noise model, which is applied post-state-preparation. We let each qubit be subject to a depolarizing channel of the form $\mathcal{E}(\hat\rho) = (1-p_\mathrm{error}) \hat\rho + p_\mathrm{error}\hat{I}/2$, where $p$ is the error rate. When any $n-1$ qubits are measured, the remaining qubit will be depolarized if \emph{any} of the original qubits were depolarized. The effective error rate on the remaining qubit is therefore $p_\mathrm{eff}=1-(1-p_\mathrm{error})^n$. In the regime of small $p_\mathrm{error}$, $p_\mathrm{eff}$ scales roughly linearly with $n$. For larger $p_\mathrm{error}$, $p_\mathrm{eff}$ exhibits asymptotic behavior, approaching $1$ for large $n$.

We now turn our attention to our first example, a scheme for implementing scalable quantum computing using probabilistic entangling gates \cite{bib:BarrettKok05,bib:Duan05}. This scheme is very general and applicable to architectures where there is negligible qubit loss, but entangling operations have non-zero failure probability ($1-p_\mathrm{gate}$). By exploiting the properties of cluster states it is shown that scalable quantum computing is possible for \emph{any} non-zero $p_\mathrm{gate}$, at the expense of a polynomial resource overhead.

\begin{figure}[!htb]
\includegraphics[width=\columnwidth]{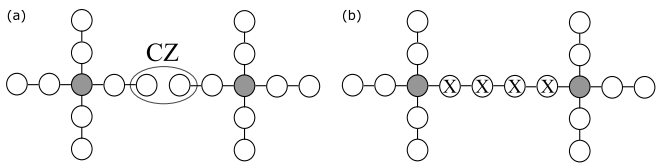}
\caption{Gate-failure-tolerant approach to constructing cluster states. The fundamental building block is the `+'-cluster. This has a central node (shown in gray) which will ultimately belong to the constructed square lattice. The central node is bonded to four linear chain clusters, each of length $n_l$. These `arms' provide redundancy, allowing multiple bonding attempts. To grow a cluster, rather than bond two cluster qubits together directly, we utilize +-clusters and attempt bonding starting at the ends of the arms (a). If this fails we lose two qubits from the respective arms, but can recover the remainder of the cluster by measuring the neighboring qubits in the $Z$-eigenbasis. We can keep reattempting the gate until there are no qubits remaining in the arms. When bonding succeeds we have the two desired cluster nodes with some remaining arm qubits left between them. These are removed by measuring them in the $X$-eigenbasis (b).} \label{fig:failure}
\end{figure}

The scheme describes how to efficiently construct square lattice cluster states. Efficient scaling is achieved using a resource of `+'-clusters. These have `arms' of redundant qubits which allow multiple attempts at applying entangling gates. The protocol is described in Fig.~\ref{fig:failure}.

After successfully bonding two arms we are left with an irregular lattice cluster which contains leftover arm qubits. The final step in the protocol is to reduce the cluster to a square lattice by measuring out these leftover qubits. This reduction stage is very similar to the multi-qubit error teleportation scenario described previously \footnote{Cluster states are \emph{not} maximally entangled. However here they exhibit similar error teleportation characteristics.}. Consider the case where the first bonding attempt between two +-clusters succeeds. We have two central node qubits with $2n_l$ redundant arm qubits remaining between them, which must all be measured out. Whenever one of these qubits suffers a phase-flip an error will be teleported onto the root qubit. Following reduction of all redundant arm qubits, an error will remain on the prepared cluster qubit if an odd number of $Z$-errors occured on the measured arm qubits. This probability scales exponentially with $n_l$, which is inversely proportional to $p_\mathrm{gate}$. Therefore, for a given effective error rate, the tolerable physical error rate scales down exponentially with $p_\mathrm{gate}$. Thus, while the scheme can tolerate arbitrary $p_\mathrm{gate}$ in principle, in practise it is fundamentally limited.


Other related proposals, such as Nielsen's \cite{bib:Nielsen04} \emph{micro-cluster} approach to efficiently constructing cluster states using non-deterministic gates, ought to exhibit similar characteristics since they also rely on reducing clusters by measuring out redundant qubits. Both these schemes are variations of the `divide-and-conquer' approach to state preparation. This is a common trick to overcoming exponential reduction in success probability in the presence of loss or gate failure and has applications beyond quantum computing. Our results potentially have broad implications for state preparation protocols. For example, Kieling \emph{et al.} \cite{bib:Kieling06} recently investigated optimal strategies for constructing cluster states using non-deterministic gates. Their analysis was entirely classical, and attempted to optimize physical resource requirements. Our results suggest that such analyses ought to be re-evaluated to consider error propagation properties.

Let us substantiate this further by considering a simple comparison of two state preparation strategies: a `single-shot', and a divide-and-conquer approach. Divide-and-conquer is clearly superior from a physical resource perspective since it exhibits polynomial resource scaling compared to the exponential scaling of the single-shot approach. However, from a fault-tolerance perspective things are quite different. Divide-and-conquer necessarily requires the reduction of redundant qubits, which propagates errors. Single-shot on the other hand does not. In this simple comparison it is evident that resource and error scaling are competing parameters.

\begin{figure}[!htb]
\includegraphics[width=0.9\columnwidth]{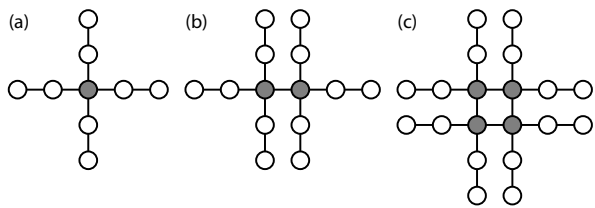}
\caption{Examples of different resource states that can be employed in the scalable construction of cluster states using non-deterministic gates.} \label{fig:resource}
\end{figure}

This observation suggests an approach for minimizing error accumulation effects in divide-and-conquer based approaches. Consider the protocol discussed previously. Referring to Fig.~\ref{fig:resource}, we begin with a resource of clusters of the form shown in (a), which we fuse together to form clusters of form (b). Similarly, two (b) clusters can be used to construct a cluster of form (c). Suppose the initial resource of +-clusters is produced using a single-shot approach. Thus, the initial resource states do not suffer from accumulated errors. Ordinarily a (b) cluster suffers error accumulation associated with the measurement of redundant arm qubits. This can obviously be avoided by instead beginning with a resource of (b) clusters, prepared using a single-shot approach. This avoids the measurement of the interstitial redundant qubits. In general, error accumulation can be further suppressed by beginning with larger resource states.

This technique effectively allows us to tailor a strategy which presents an arbitrary tradeoff between the single-shot and divide-and-conquer strategies. The tradeoff between competing resources is clear. For a given bound on the effective error rate, using larger resource states allows us to tolerate higher local error rates, since error accumulation is reduced. However, because they are prepared using a single-shot approach, this requires physical resources growing exponentially with their size, and polynomially with gate failure rate. This places fundamental limitations on practically tolerable gate failure rates.

A simple numerical example is illustrative. From Ref.~\cite{bib:Duan05}, constructing a 100 qubit cluster state with 10\% success probability, using {\sc CPHASE} gates operating with 99\% success probability requires a resource of +-clusters with arm length $n_l\approx 11$. Suppose we construct the resource states using single-shot. The preparation of each +-cluster succeeds with probability $p_\mathrm{success}={p_\mathrm{gate}}^{4n_l}\approx 0.64$. Next we join two +-clusters together to form a cluster of type (b). With a physical depolarizing rate of $p_\mathrm{error}=10^{-3}$, after measurement of redundant qubits, the effective depolarizing rate is $p_\mathrm{eff}\approx 1.1\times 10^{-2}$, an order of magnitude increase. Alternately, we could produce type-(b) clusters directly. Now the single-shot success probability is $p_\mathrm{success}={p_\mathrm{gate}}^{6n_l+1}\approx 0.51$. However, there are no accumulated errors associated with joining the +-clusters, so the effective error rate is just the physical error rate of $10^{-3}$.

While this example exhibits a significant reduction in effective error rates, it is clear that we are limited to a high $p_\mathrm{gate}$ regime. For lower values of $p_\mathrm{gate}$, we loose our ability to directly prepare type-(b) clusters, and single-shot can only be used to construct smaller states. While this approach is limited, this example illustrates the benefits of shifting as much of state preparation into single-shot as possible.


As a second example we consider the Varnava \emph{et al.} \cite{bib:Varnava05} approach to tolerating qubit loss in cluster states. This scheme relies on the principle of \emph{indirect measurement}, where the measurement outcome of a lost qubit can be inferred by measuring correlated qubits.

The important feature of cluster states, from which indirect measurement properties follow, is their stabilizer representation. Associated with every qubit $i$ in a cluster state is a stabilizer of the form $\hat{S}_i=\hat{X}_i\bigotimes_{j\in v(i)}\hat{Z}_j$, where $v(i)$ is the set of qubits neighboring $i$. The stabilizers define correlations in measurement outcomes. Indirect measurement exploits these correlations to infer the measurement result of a lost qubit using only the measurement results of correlated qubits.

\begin{figure}[!htb]
\includegraphics[width=0.5\columnwidth]{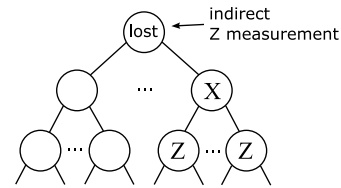}
\caption{Using a tree cluster to perform indirect $Z$-measurement of a lost qubit. The qubit below the lost qubit is measured in the $X$-eigenbasis, and each of the qubits below that in the $Z$-eigenbasis. If the $X$-measurement fails, we can make another attempt on the next branch. If any of the $Z$-measurements fail they can be indirectly measured by moving further down the tree.} \label{fig:cluster}
\end{figure}

In this scheme each cluster qubit is replaced with a `tree' cluster, with its root node planted in place of the cluster qubit. The tree structure facilitates multiple attempts at indirect measurement of a lost root qubit, suppressing effective loss rates. The is described in Fig.~\ref{fig:cluster}.

Indirect measurement exhibits similar error teleportation properties to the previous example -- an error will propagate onto the lost root qubit if an odd number of measurement results were incorrect. Based on results from Ref.~\cite{bib:Varnava05}, achieving an effective loss rate of $\varepsilon_\mathrm{eff}\approx 10^{-3}$ given a physical loss rate of $\varepsilon_\mathrm{loss}=0.2$, requires tree clusters with roughly $Q\approx 1000$ qubits. Suppose an indirect measurement requires measuring half the tree on average. This will magnify a physical error rate of $p_\mathrm{error}\approx 10^{-3}$ to an effective error rate on the indirectly measured qubit of $p_\mathrm{eff}\approx 0.32$, an increase of more than two orders of magnitude.

This scheme can also be modified to overcome exponential error scaling through a minor adjustment to the protocol. Referring to Fig.~\ref{fig:cluster}, we have multiple attempts at a given indirect measurement, one for each branch in the tree. While in principle only one indirect measurement is required to measure a lost qubit, by utilizing all available branches we can exploit the fact that all indirect measurement outcomes ought to be consistent and implement a majority vote. This was first recognized by Browne, Rudolph and Varnava \cite{bib:BrowneRudolphVarnava05}. If indirect measurement is performed in parallel via $b_1$ \footnote{$b_1\dots b_d$ denote the tree's branching parameters -- the number of branches that emanate from each node at the respective level of the tree. $d$ is the depth of the tree.} branches, the probability of an error propagating into the measurement outcome scales as $p_\mathrm{eff}=\textsc{Exp}^{-1}(b_1)$ with $p_\mathrm{single}$, the probability of any single indirect measurement being incorrect. On the other hand, $p_\mathrm{single}$ scales as $p_\mathrm{single}=\textsc{Exp}(\textsc{Poly}[b_2,\dots,b_d])$ with $p_\mathrm{error}$, where $\textsc{Poly}$ represents some polynomial function of its parameters. Therefore, for an appropriate choice of branching parameters $\{b_i\}$, one expects that exponential error scaling can be eliminated.

Loss rates determine the \emph{effective} value of $b_1$. Thus, higher loss rates imply lower confidence in the majority vote, increasing error rates. This undermines the otherwise very high loss thresholds promised by this scheme. To illustrate this, we performed a numerical analysis of a simple two-level tree structure with branching parameters $b_1=b_2=3$. This structure improves the effective loss rate (i.e. $\varepsilon_\mathrm{eff}\leq\varepsilon_\mathrm{loss}$) for $\varepsilon_\mathrm{loss}\lesssim 0.195$. Under the original scheme, this loss rate would increase an error rate of $p_\mathrm{error}=10^{-3}$ to an effective error rate of $p_\mathrm{eff}\approx 4\times 10^{-3}$. With the introduction of majority voting this reduces to $p_\mathrm{eff}\approx 1.7\times 10^{-3}$. Furthermore, there is a `break-even' point on $\varepsilon_\mathrm{loss}$, below which there is no degradation in error rates (i.e. $p_\mathrm{eff}\leq p_\mathrm{error}$). In this example this occurs at $\varepsilon_\mathrm{loss}\approx 0.1$, roughly half the in-principle loss tolerance rate. Thus, if the scheme is to be operated in a regime where error rates do not suffer, the loss threshold is significantly reduced. 

Other loss-tolerant architectures ought to exhibit similar properties to those presented here for loss-tolerant cluster states. For example, in the loss-tolerant scheme of Ralph \emph{et al.} \cite{bib:RalphHayes05} logical qubits are encoded into maximally entangled \emph{parity states}. One of the fundamental operations in this scheme is re-encoding, where new qubits are `grafted' onto an existing parity state and all the old ones measured out. This provides a situation completely analogous to the multi-qubit teleportation scenario and exhibits identical exponential error scaling properties.

Recent proposals for loss- and failure-tolerance have been constructed to deal with a specific and very limited error model. Consequently, they often promise extremely high loss/failure thresholds. While effective within this limited context their disadvantage is that the effects of other error types are magnified exponentially. This arises from the introduction of redundant qubits, which provide new opportunities for errors to occur. This fundamentally limits the extent to which such schemes can be used and undermines their loss/failure tolerance.

For the examples cited we discussed techniques to minimize the error scaling problem. In the case of the failure-tolerant scheme, while beneficial in terms of noise tolerance, the discussed solution presented a direct tradeoff against physical and temporal resource requirements. The solution for the loss-tolerant scheme resulted in significantly reduced loss thresholds. Our results suggest that, despite their limitations, these schemes may be useful for dealing with \emph{modest} loss rates.


An important point is that our discussion is in a non-fault-tolerant context. We assume a resource of \emph{perfect} resource states and that noise acts locally on these states after construction. In practise, state preparation introduces a plethora of new opportunities for errors to be introduced, including correlated errors, which none of the presented  solutions can deal with effectively. Such effects will further reduce the loss/failure tolerance of these schemes. A comparison with fault tolerant schemes is illustrative. Fault tolerant thresholds for joint depolarizing and photon loss errors in the cluster state model were recently studied by Dawson \emph{et al.} \cite{bib:Dawson05}. The loss threshold was estimated to be on the order of $3\times 10^{-3}$, two orders of magnitude less than that achievable using the loss-tolerant schemes discussed here.

While we have demonstrated the concept of error scaling by example of several well-known protocols, we believe our results have broad implications for loss- and failure-tolerant protocols, state preparation strategies, and potentially other schemes which make use of redundant encoding or ancillary states. The central message is the following: any fault-tolerant protocol must tolerate a \emph{general} class of errors. Inevitably, codes tailored to a specific error type will be more sensitive to others. Significant work has been done into developing codes protecting against depolarizing noise \cite{bib:CalderbankShor96,bib:Shor95,bib:Steane96,bib:NielsenChuang00}. One might question whether such codes suffer because they do not explicitly accommodate for loss errors. This is not the case since qubit loss can always be trivially mapped to a depolarizing error, because qubit loss is a located error. However depolarizing errors cannot be mapped to loss, because they are unlocated \cite{bib:Rohde06}. It is this distinction that makes considerations in the construction of loss-specific codes inherently different from depolarization specific codes.

A detailed analysis of the discussed solutions to error scaling will be presented in a supplementary paper.

\begin{acknowledgments}
We thank Michael Nielsen for the discussion that motivated this work, and Henry Haselgrove and Alex Hayes for helpful discussions. This work was supported by the Australian Research Council and QLD State Government. We acknowledge partial support by the DTO-funded U.S. Army Research Office Contract No. W911NF-05-0397.
\end{acknowledgments}

\bibliography{paper}

\begin{thebibliography}{18}
\expandafter\ifx\csname natexlab\endcsname\relax\def\natexlab#1{#1}\fi
\expandafter\ifx\csname bibnamefont\endcsname\relax
  \def\bibnamefont#1{#1}\fi
\expandafter\ifx\csname bibfnamefont\endcsname\relax
  \def\bibfnamefont#1{#1}\fi
\expandafter\ifx\csname citenamefont\endcsname\relax
  \def\citenamefont#1{#1}\fi
\expandafter\ifx\csname url\endcsname\relax
  \def\url#1{\texttt{#1}}\fi
\expandafter\ifx\csname urlprefix\endcsname\relax\def\urlprefix{URL }\fi
\providecommand{\bibinfo}[2]{#2}
\providecommand{\eprint}[2][]{\url{#2}}

\bibitem[{\citenamefont{Knill et~al.}(2001)\citenamefont{Knill, Laflamme, and
  Milburn}}]{bib:KLM01}
\bibinfo{author}{\bibfnamefont{E.}~\bibnamefont{Knill}},
  \bibinfo{author}{\bibfnamefont{R.}~\bibnamefont{Laflamme}}, \bibnamefont{and}
  \bibinfo{author}{\bibfnamefont{G.}~\bibnamefont{Milburn}},
  \bibinfo{journal}{Nature (London)} \textbf{\bibinfo{volume}{409}},
  \bibinfo{pages}{46} (\bibinfo{year}{2001}).

\bibitem[{\citenamefont{Kok et~al.}(2005)\citenamefont{Kok, Munro, Ralph,
  Dowling, and Milburn}}]{bib:Kok05}
\bibinfo{author}{\bibfnamefont{P.}~\bibnamefont{Kok}},
  \bibinfo{author}{\bibfnamefont{W.~J.} \bibnamefont{Munro}},
  \bibinfo{author}{\bibfnamefont{T.~C.} \bibnamefont{Ralph}},
  \bibinfo{author}{\bibfnamefont{J.~P.} \bibnamefont{Dowling}},
  \bibnamefont{and} \bibinfo{author}{\bibfnamefont{G.~J.}
  \bibnamefont{Milburn}} (\bibinfo{year}{2005}), \eprint{quant-ph/0512071}.

\bibitem[{\citenamefont{Varnava et~al.}(2005)\citenamefont{Varnava, Browne, and
  Rudolph}}]{bib:Varnava05}
\bibinfo{author}{\bibfnamefont{M.}~\bibnamefont{Varnava}},
  \bibinfo{author}{\bibfnamefont{D.~E.} \bibnamefont{Browne}},
  \bibnamefont{and} \bibinfo{author}{\bibfnamefont{T.}~\bibnamefont{Rudolph}}
  (\bibinfo{year}{2005}), \eprint{quant-ph/0507036}.

\bibitem[{\citenamefont{Ralph et~al.}(2005)\citenamefont{Ralph, Hayes, and
  Gilchrist}}]{bib:RalphHayes05}
\bibinfo{author}{\bibfnamefont{T.~C.} \bibnamefont{Ralph}},
  \bibinfo{author}{\bibfnamefont{A.~J.~F.} \bibnamefont{Hayes}},
  \bibnamefont{and}
  \bibinfo{author}{\bibfnamefont{A.}~\bibnamefont{Gilchrist}},
  \bibinfo{journal}{Phys. Rev. Lett.} \textbf{\bibinfo{volume}{95}},
  \bibinfo{pages}{100501} (\bibinfo{year}{2005}).

\bibitem[{\citenamefont{Barrett and Kok}(2005)}]{bib:BarrettKok05}
\bibinfo{author}{\bibfnamefont{S.~D.} \bibnamefont{Barrett}} \bibnamefont{and}
  \bibinfo{author}{\bibfnamefont{P.}~\bibnamefont{Kok}},
  \bibinfo{journal}{Phys. Rev. A} \textbf{\bibinfo{volume}{71}},
  \bibinfo{pages}{060310(R)} (\bibinfo{year}{2005}).

\bibitem[{\citenamefont{Duan and Raussendorf}(2005)}]{bib:Duan05}
\bibinfo{author}{\bibfnamefont{L.-M.} \bibnamefont{Duan}} \bibnamefont{and}
  \bibinfo{author}{\bibfnamefont{R.}~\bibnamefont{Raussendorf}},
  \bibinfo{journal}{Phys. Rev. Lett.} \textbf{\bibinfo{volume}{95}},
  \bibinfo{pages}{080503} (\bibinfo{year}{2005}).

\bibitem[{\citenamefont{Nielsen}(2006)}]{bib:Nielsen06}
\bibinfo{author}{\bibfnamefont{M.~A.} \bibnamefont{Nielsen}},
  \bibinfo{journal}{Rep. Math. Phys.} \textbf{\bibinfo{volume}{57}},
  \bibinfo{pages}{147} (\bibinfo{year}{2006}).

\bibitem[{\citenamefont{Raussendorf and Briegel}(2001)}]{bib:Raussendorf01}
\bibinfo{author}{\bibfnamefont{R.}~\bibnamefont{Raussendorf}} \bibnamefont{and}
  \bibinfo{author}{\bibfnamefont{H.~J.} \bibnamefont{Briegel}},
  \bibinfo{journal}{Phys. Rev. Lett.} \textbf{\bibinfo{volume}{86}},
  \bibinfo{pages}{5188} (\bibinfo{year}{2001}).

\bibitem[{\citenamefont{Raussendorf et~al.}(2003)\citenamefont{Raussendorf,
  Browne, and Briegel}}]{bib:Raussendorf03}
\bibinfo{author}{\bibfnamefont{R.}~\bibnamefont{Raussendorf}},
  \bibinfo{author}{\bibfnamefont{D.~E.} \bibnamefont{Browne}},
  \bibnamefont{and} \bibinfo{author}{\bibfnamefont{H.~J.}
  \bibnamefont{Briegel}}, \bibinfo{journal}{Phys. Rev. A}
  \textbf{\bibinfo{volume}{68}}, \bibinfo{pages}{022312}
  (\bibinfo{year}{2003}).

\bibitem[{\citenamefont{Nielsen}(2004)}]{bib:Nielsen04}
\bibinfo{author}{\bibfnamefont{M.~A.} \bibnamefont{Nielsen}},
  \bibinfo{journal}{Phys. Rev. Lett.} \textbf{\bibinfo{volume}{93}},
  \bibinfo{pages}{040503} (\bibinfo{year}{2004}).

\bibitem[{\citenamefont{Kieling et~al.}(2006)\citenamefont{Kieling, Gross, and
  Eisert}}]{bib:Kieling06}
\bibinfo{author}{\bibfnamefont{K.}~\bibnamefont{Kieling}},
  \bibinfo{author}{\bibfnamefont{D.}~\bibnamefont{Gross}}, \bibnamefont{and}
  \bibinfo{author}{\bibfnamefont{J.}~\bibnamefont{Eisert}}
  (\bibinfo{year}{2006}), \eprint{quant-ph/0601190}.

\bibitem[{\citenamefont{Browne et~al.}(2005)\citenamefont{Browne, Rudolph, and
  Varnava}}]{bib:BrowneRudolphVarnava05}
\bibinfo{author}{\bibfnamefont{D.~E.} \bibnamefont{Browne}},
  \bibinfo{author}{\bibfnamefont{T.}~\bibnamefont{Rudolph}}, \bibnamefont{and}
  \bibinfo{author}{\bibfnamefont{M.}~\bibnamefont{Varnava}}
  (\bibinfo{year}{2005}).

\bibitem[{\citenamefont{Dawson et~al.}(2005)\citenamefont{Dawson, Haselgrove,
  and Nielsen}}]{bib:Dawson05}
\bibinfo{author}{\bibfnamefont{C.~M.} \bibnamefont{Dawson}},
  \bibinfo{author}{\bibfnamefont{H.~L.} \bibnamefont{Haselgrove}},
  \bibnamefont{and} \bibinfo{author}{\bibfnamefont{M.~A.}
  \bibnamefont{Nielsen}}, \bibinfo{journal}{Phys. Rev. Lett.}
  \textbf{\bibinfo{volume}{96}}, \bibinfo{pages}{020501}
  (\bibinfo{year}{2005}).

\bibitem[{\citenamefont{Calderbank and Shor}(96)}]{bib:CalderbankShor96}
\bibinfo{author}{\bibfnamefont{A.~R.} \bibnamefont{Calderbank}}
  \bibnamefont{and} \bibinfo{author}{\bibfnamefont{P.~W.} \bibnamefont{Shor}},
  \bibinfo{journal}{Phys. Rev. A} \textbf{\bibinfo{volume}{54}},
  \bibinfo{pages}{1098} (\bibinfo{year}{96}).

\bibitem[{\citenamefont{Shor}(1995)}]{bib:Shor95}
\bibinfo{author}{\bibfnamefont{P.~W.} \bibnamefont{Shor}},
  \bibinfo{journal}{Phys. Rev. A} \textbf{\bibinfo{volume}{52}},
  \bibinfo{pages}{R2493} (\bibinfo{year}{1995}).

\bibitem[{\citenamefont{Steane}(1996)}]{bib:Steane96}
\bibinfo{author}{\bibfnamefont{A.~M.} \bibnamefont{Steane}},
  \bibinfo{journal}{Phys. Rev. Lett.} \textbf{\bibinfo{volume}{77}},
  \bibinfo{pages}{793} (\bibinfo{year}{1996}).

\bibitem[{\citenamefont{Nielsen and Chuang}(2000)}]{bib:NielsenChuang00}
\bibinfo{author}{\bibfnamefont{M.~A.} \bibnamefont{Nielsen}} \bibnamefont{and}
  \bibinfo{author}{\bibfnamefont{I.~L.} \bibnamefont{Chuang}},
  \emph{\bibinfo{title}{Quantum Computation and Quantum Information}}
  (\bibinfo{publisher}{Cambridge University Press, Cambridge},
  \bibinfo{year}{2000}).

\bibitem[{\citenamefont{Rohde}(2006)}]{bib:Rohde06}
\bibinfo{author}{\bibfnamefont{P.~P.} \bibnamefont{Rohde}}
  (\bibinfo{year}{2006}), \eprint{quant-ph/0605183}.

\end{thebibliography}

\end{document}